\documentclass[12pt,a4paper]{article}
\usepackage{amsmath}
\usepackage{amsthm}
\usepackage{amsfonts}
\usepackage{multicol}
\usepackage[dvips]{graphicx}
\usepackage{fixltx2e}

\newcommand{\danger}[1]{\textbf{#1}}

\input{epsf}             % encapsulated Postscript figures
\begin{document}

\title{\danger{Entropic Motion in Loop Quantum Gravity}}
\author{\centerline{\danger{J. Manuel Garc\'\i a-Islas \footnote{
e-mail: jmgislas@iimas.unam.mx}}}  \\
Departamento de F\'\i sica Matem\'atica \\
Instituto de Investigaciones en Matem\'aticas Aplicadas y en Sistemas \\ 
Universidad Nacional Aut\'onoma de M\'exico, UNAM \\
A. Postal 20-726, 01000, M\'exico DF, M\'exico\\}

\maketitle

\begin{abstract}
Entropic forces result from an increase of the entropy of a thermodynamical physical system. It has been
proposed that gravity is such a phenomenon and many articles have appeared on the literature concerning
this problem. Loop quantum gravity has also considered such possibility. We propose a new method
in loop quantum gravity which reproduces an entropic force. 
By considering the interaction
between a fixed gravity state space and a particle state in loop quantum gravity, we show that it 
leads to a mathematical description of a random walk of such particle. The random walk in special situations,
can be seen as an entropic motion in such a way that the particle will move towards a location where entropy increases. 
This may prove that such theory can
reproduce gravity as it is expected.   
\end{abstract}

\section{Introduction}

Loop quantum gravity \cite{r}, \cite{t} is a theory of quantum geometry. 
It describes the quantisation of general relativity, and as such
it is expected to reproduce gravity in a semiclassical limit \cite{t2}, \cite{t3}, \cite{ac}.  
On the other hand, gravity has been described as an entropic force \cite{v} and this idea 
has been studied in many different fields including
loop quantum gravity \cite{s}, \cite{k}.

In this article we propose a way to produce an entropic force in loop quantum gravity which involves
the mathematical technology of random walks on graphs. 
We will work on this in the next section. We start by considering the interaction of a gravity state space
and a particle state space as described in \cite{rv}.

Our proposal then is that the Hamiltonian of the particle on a fixed gravity state space can be thought of as
describing a random walk on a special weighted graph. The random walk is a motion on the graph,
in a probabilistic way, such that there is a tendency for the particle to be localised most probable 
at some vertices than at others. 

By defining a local entropy at each vertex of the graph, at special cases we can make the particle to be
localised at the most probable vertex and at the place where entropy is maximised. 
This clearly points at some kind of entropic force which can be interpreted as a result of the
interaction of a particle with some special gravity state space.  Or in other words, loop quantum gravity
is really able to reproduce gravity as we know it.  

Could this special gravity state spaces
be semiclassical? Could they guide us to where we should look in order to obtain
the semiclassical limit of loop quantum gravity?

\section{Entropic motion of a particle}

In this section we describe our proposal of entropic motion in loop quantum gravity which may lead us to
entropic forces  and to gravity. Throughout this article all physical constants used in loop quantum
gravity are considered to be equal to $1$; except for $G$, the Newton's gravitational constant.

\bigskip

The quantum states of the gravitational field in loop quantum gravity, are given by a set of graphs, called spin networks.\footnote{
A graph is a pair $\Gamma=(V(\Gamma), E(\Gamma))$
of finite(or countable) sets, such that $V(\Gamma) \neq \emptyset$ and $E(\Gamma)$ consist of unordered pairs of elements of
$V(\Gamma)$. The set $V(\Gamma)$ is known as the set of vertices, and the set $E(\Gamma)$ is known as the
set of edges.} 
Each spin network graph $\Gamma=(V(\Gamma), E(\Gamma))$, has an edge colouring and a vertex colouring.\footnote{
In this paper, we will only be concerned with the edge colouring.}
An edge colouring
is a function $c: E(\Gamma) \rightarrow I$, where $I=\{ 1/2 , 1 , 3/2 , 2 ,... \}$ is the set of positive
half integers.\footnote{Edge colouring in the mathematical literature has the additional condition,
$c(e_i) \neq c(e_j)$ for any adjacent edges.} It is customary to denote the elements of $V(\Gamma)$ by $v_1 , v_2 , ...,v_k ,...$,
and the elements of $E(\Gamma)$ by $e_1 , e_2 , ... e_k,..$,
as well as the colouring, $c(e_k) = j_k$, (Figure 1).\footnote{The shadowed face is only for aesthetic purposes. Graphs
have only vertices and edges.}

\begin{figure}[h]
\begin{center}
\includegraphics[width=0.6\textwidth]{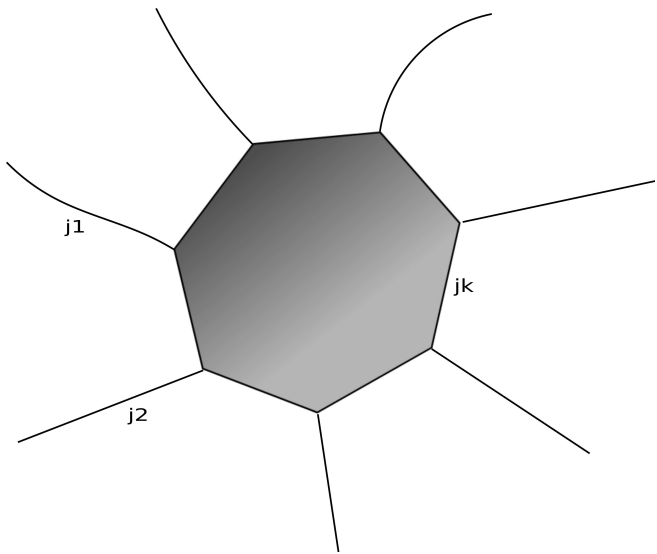}
\caption{A spin network representing a quantum space}
\end{center}
\end{figure}
We will use a different notation for the edges $E(\Gamma)$. If vertex $v_m$ is joined by an edge to vertex $v_n$,
we denote it by $e_{mn}$ and its colour by $c(e_{mn}) = j_{mn}$. 

The set of vertices are thought of as chunks of space containing quantum volume whereas the set of edges are
thought of as quantum surfaces separating chunks of space.  
The colour of edge $e_{mn}$, represents a quantum surface which area is discrete and given by 

\begin{equation}
A(j_{mn})=  \sqrt{j_{mn}(j_{mn}+1)}
\end{equation}
A spin network graph $\Gamma$ is interpreted in loop quantum gravity, as a discrete quantum space.  

\bigskip

Let $\bar{\Gamma}$ be a graph which vertices and edges are the same as those of 
the spin network $\Gamma$, but 
weighted in a different way.
Let the colour of edge $e_{mn}$ be $c(e_{mn}) = j_{mn} (j_{mn} +1)$, see Figure [2]. Let us focus on graph $\bar{\Gamma}$
and consider some graph theory definitions \cite{bm}, \cite{bm2}.

Define the degree of a vertex $v_m \in V(\bar{\Gamma})$ by $d_m= \sum_{n} j_{mn} (j_{mn}+1)$, where
the sum is over all edges adjacent to $v_m$.

The adjacency matrix of weighted graph $\bar{\Gamma}$, is an $n \times n$ matrix defined by
$\mathcal{A}(\bar{\Gamma}) = ( j_{mn}(j_{mn} + 1) )$, 
 with rows and columns indexed by $V(\bar{\Gamma})$, and
whose entries are the edge weights.

The degree matrix of weighted graph $\bar{\Gamma}$, is defined by 
$\mathcal{D}(\bar{\Gamma}) = \text{diag} (d_m)$. It 
is the diagonal matrix indexed by $V(\bar{\Gamma})$ with the degree of the vertices on the diagonal.

Define the matrix

\begin{equation}
\mathcal{L}(\bar{\Gamma}) = \mathcal{D}(\bar{\Gamma})  - \mathcal{A}(\bar{\Gamma}) 
\end{equation}
Consider the set of all functions $\psi : V \longrightarrow \mathbf{R}$ from the set of vertices
to the real numbers. This set of functions defines a vector space isomorphic to $\mathbf{R}^{\mid V \mid}$
endowed with the inner product

\begin{equation}
< \psi , \phi > \ = \sum_{v_m \in V} \psi(v_m) \phi(v_m)
\end{equation}

\begin{figure}[h]
\begin{center}
\includegraphics[width=0.6\textwidth]{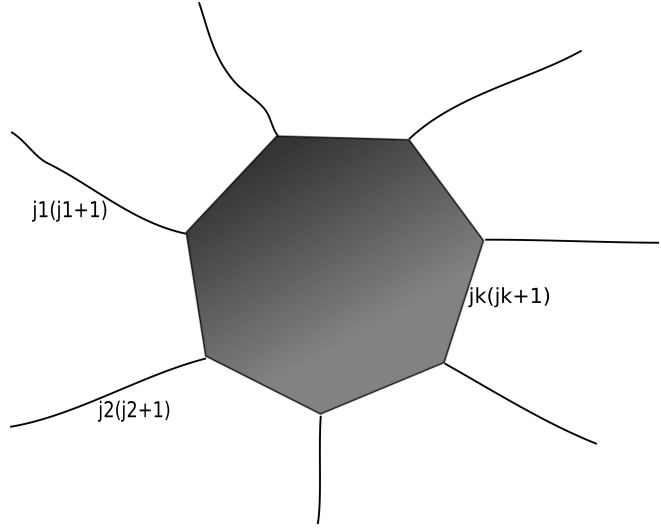}
\caption{Graph $\bar{\Gamma}$ associated to the particle}
\end{center}
\end{figure}

The matrix $\mathcal{L}(\bar{\Gamma})$ can be thought of as an operator acting on $\mathbf{R}^{\mid V \mid}$
by matrix and vector multiplication.

It is not difficult to see that

\begin{equation}
< \psi , \mathcal{L}(\bar{\Gamma}) \psi > \ = \sum_{e} \ j_{mn} (j_{mn} +1) \ [ \psi(v_m) - \psi(v_n) ]^2
\end{equation}

Up to a constant, which we consider equal to one, formula $(4)$ is identical to 
the Hamiltonian operator $H$ of a particle
on a gravitational field introduced in \cite{rv}. Therefore, we can denote 

\begin{equation}
<\psi \mid H \mid \psi> = \ < \psi , \mathcal{L}(\bar{\Gamma}) \psi > 
\end{equation}
However, we have described a completely combinatorial way to arrive at formula $(4)$. 
It was done by considering 
graph theory concepts. 
 
It is natural to consider the random motion of a particle on the graph $\bar{\Gamma}$. The reason for
this is that the matrix $\mathcal{L}(\bar{\Gamma})$ is related to this motion.

Define a random walk of a particle on $\bar{\Gamma}$ as follows:
a single particle located at vertex $v_m$ will move to 
a neighbour vertex $v_n$ with probability

\begin{equation}
P(v_m, v_n)= \frac{j_{mn} (j_{mn}+1)}{\sum_{k} j_{mk} (j_{mk}+1)} =
\frac{j_{mn} (j_{mn}+1)}{d_m} 
\end{equation}
Where $j_{mn} (j_{mn}+1)$ is the colour of the edge which joins vertex $v_m$ to vertex $v_n$,
and the sum on the denominator is over all edges incident to $v_m$.

This random walk is known by simple random walk on $\bar{\Gamma}$.

It can be seen that 

\begin{equation}
\sum_{n} P(v_m, v_n)= 1
\end{equation}
The probabilities $P(v_m, v_n)$ are called transition probabilities.

We can associate a matrix $P(\bar{\Gamma})$
to these given probabilities indexed by $V(\bar{\Gamma})$. This matrix is known as 
the transition matrix for this simple random
walk. It is easy to see that this matrix satisfies

\begin{equation}
P(\bar{\Gamma}) = \mathcal{D}(\bar{\Gamma})^{-1}  \mathcal{A}(\bar{\Gamma}) 
\end{equation}
and that

\begin{equation}
P(\bar{\Gamma}) = I - \mathcal{D}(\bar{\Gamma})^{-1}  \mathcal{L}(\bar{\Gamma})
\end{equation}
where $I$ denotes the identity matrix.

In a random walk, 
the probability, $\pi(v_n)$, that the particle ends up at an arbitrary $v_n$ vertex,
after a long period of time is given by

\begin{equation}
\pi(v_n)= \frac{d_{n}}{\sum_{n} d_n} = \frac{d_{n}}{\text{Vol}(\bar{\Gamma})}
\end{equation} 
where we denote, $\text{Vol}(\bar{\Gamma}) = \sum_{n} d_n$.

These probabilities mean that there is a vertex $v_n$, where the particle is more likely to be found, whenever
 
\begin{equation}
\pi(v_n) \geq \pi(v_m)
\end{equation} 
for all the remaining vertices.

It can be seen that the way we have defined the probability transitions and the probabilities $\pi(v_n)$, it
is satisfied

\begin{eqnarray}
& \sum_{n} \pi(v_n)= 1 \nonumber \\
& \nonumber \\
& \sum_{n} \pi(v_n) P(v_n, v_m) = \pi(v_m)
\end{eqnarray} 
The first equality is trivial. For the second one, it is also easily verified since

\begin{eqnarray}
& \sum_{n} \pi(v_n) P(v_n, v_m) = \sum_{n} \frac{d_{n}}{\text{Vol}(\bar{\Gamma)}} \frac{j_{mn} (j_{mn}+1)}{d_n} = \nonumber \\
& =  \sum_{n} \frac{j_{mn} (j_{mn}+1)}{\text{Vol}(\bar{\Gamma)}} 
= \frac{d_m}{\text{Vol}(\bar{\Gamma)}} = \pi(v_m)
\end{eqnarray}  
These last equalities just tell us that the random walk we are considering is mathematically a specific class
of a Markov chain, 
see \cite{w}, \cite{o} for an introduction to Markov chains. 

Define the local entropy at each vertex $v_n$ by 

\begin{align}
S_{v_n}  & =  - \sum_{m} P(v_n , v_m) \log P(v_n , v_m) \nonumber \\
     & = - \sum_{m} \frac{j_{nm}(j_{nm}+1)}{\text{Vol}(\Gamma)} \log  \frac{j_{nm}(j_{nm}+1)}{\text{Vol}(\Gamma)} 
\end{align}
This entropy is clearly maximised at a vertex when the number of edges incident to that particular vertex is large, and when 
they have equal weights. It is minimal when the weights associated to the edges incident to that vertex
are all zero but one which could have any weight associated to it.

The fact that the particle at random motion in the graph $\bar{\Gamma}$ will tend to be localised at a vertex $v_n$  
where $\pi(v_n)$ is the largest number, does not necessarily imply that the entropy at that
vertex $S_{v_n}$ is maximised. In order to
have that property we have to consider a very particular graph case.

We will consider the simplest of all cases. Suppose the graph $\bar{\Gamma}$ is such that it contains a vertex
$v_n$ with a very large number $N$ of edges incident to it. Moreover, suppose it is the vertex of
largest valance.\footnote{The valance of a vertex in a graph is the number of edges incident to it.}

Let all the weights of the edges incident to vertex $v_n$ be equal and large.\footnote{In fact
suppose these weights are the largests of all} 
Then we have that $\pi(v_n) > \pi(v_m)$, and $S_{v_n} > S_{v_m}$,
for all $m$ which labels the remaining vertices. This is the 'entropic motion' we are defining. For particular
cases the particle will move to a place where the entropy is the largest. This can be though of as an entropic
motion. The particle will tend to increase its entropy by localising itself at the proper vertex.

In this case we have that the particle starting at any vertex of the graph $\Gamma$, will be move in the graph
and be localised most probable at vertex $v_n$. The local entropy in this case is simply given by

\begin{align}
S_{v_n} = \log (N)
\end{align}
Imagine that the large number of edges incident to vertex $v_n$ are puncturing a spherical surface, like in 
an isolated black hole description of loop quantum gravity, see Figure 4.

\begin{figure}[h]
\begin{center}
\includegraphics[width=0.6\textwidth]{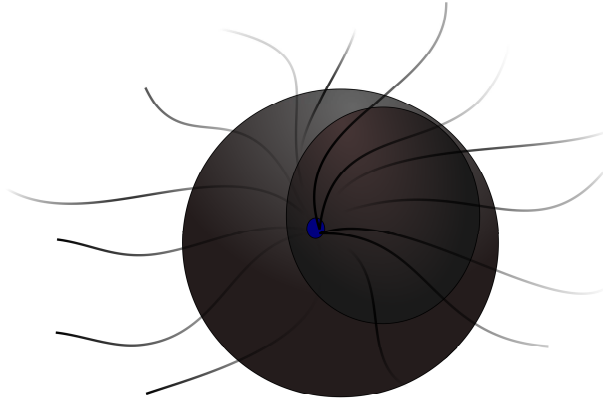}
\caption{Particle localised at a vertex of the largest valance in $\bar{\Gamma}$}
\end{center}
\end{figure}
If we follow \cite{v} and suppose that the number of punctures is related to the spherical area by

\begin{align}
N = \frac{A}{G}
\end{align}
we have that 

\begin{align}
S_{v_n} = \log (A) + C
\end{align}
where $C=\log(G)$ is a constant with no significance.
Let us mention one thing here. It is known that the entropy of a black hole of
very large horizon area in loop quantum gravity has a logarithmic correction, \cite{km}, \cite{ml}, \cite{gm},
given by $- (1/2) \ S_{v_n}$.

From this we can propose that the logarithmic correction to the entropy of a black hole in loop quantum gravity,
may be due to the interaction of quantum space with a particle. 

Now, let us consider the possibility of whether this entropic motion may reproduce gravitational entropic force
as proposed in \cite{v}.

In order to have an entropic force we need a temperature $T$, and we also need to compute the change of entropy
with respect to position. In our case as we have a discrete system given by the graph $\bar{\Gamma}$, we have
that the variation of entropy with respect to position is just a difference of the local entropies at neighbour
vertices.

This means that the change of entropy at vertex $v_n$ where the particle is moving to, is given by

\begin{align}
\frac{dS}{dx} = \mid S_{v_n} - S_{v_m} \mid
\end{align}
where $v_m$ is a neighbour vertex of $v_m$. If we suppose that vertex $v_m$ has also a very large number of edges 
$X$ incident to it, such that $X$ is close to $N$, and such that 
they have also equal weights, we will have that the difference $(18)$ is small. Following again \cite{v},
we identify this small number with
the mass $m$ of the particle

\begin{align}
\mid S_{v_n} - S_{v_m} \mid = \log \bigg( \frac{N}{X} \bigg) = m
\end{align}
As the particle is most probably moving from vertex $v_m$ to vertex $v_n$ it must cross the imaginary 
spherical surface. Again as in \cite{v}, we suppose the temperature of the surface is given by

\begin{align}
T \sim \frac{GM}{R^2}
\end{align} 
which implies that

\begin{align}
F = T \mid S_{v_n} - S_{v_m} \mid = \frac{GMm}{R^2}
\end{align}
which may imply that the entropic motion we are defining in loop quantum gravity, reproduces Newtonian gravity,
as suggested in \cite{s} also.
    
\section{Conclusions}

We have presented an idea of entropic motion and entropic force in loop quantum gravity. A toy model which could be
strengthened and become an important issue in loop quantum gravity. Any theory of quantum gravity is expected to
reproduce general relativity in a semiclassical limit and of course Newtonian Gravity. 

Loop quantum gravity has had difficulty when considering the semiclassical limit. However, important advances
have been developed, \cite{t2}, \cite{t3}, \cite{ac}. We expect to contribute in a fruitful way
with the developed idea of this paper. For instance, we have seen that given a quantum gravity state,
that is, a spin network $\Gamma$, we can associate to it a new graph $\bar{\Gamma}$, 
representing a place where a particle moves
in a random way. Given a particle which moves randomly in the graph $\bar{\Gamma}$, it happens that there will be some
places more visited than others from equation $(5)$. And we have mentioned also, what the properties of
these graphs should be in order for the random motion to be directed towards maximal local entropy.

Formula $(19)$ represents how the entropy varies according to position. This variation is proportional to
the mass of the particle, according to \cite{v}. However, in the model we present, this variation may change
from point to point and may imply a change in mass. If we want the mass to be preserved,
we should propose formula $(19)$ to remain constant throughout the motion of the particle in the
graph $\bar{\Gamma}$.

What types of graphs $\bar{\Gamma}$, are the ones for which the random motion is directed towards 
an entropic one, that is, where entropy is maximised? Moreover, what should these graphs
satisfy in order for equation $(19)$ remain constant throughout the motion of the particle?

Does it happen that the associated
spin networks $\Gamma$ represent semiclassical states? 

We propose to investigate in this latter question. It may happen that the types of spin networks $\Gamma$,
which are associated to entropic motion in $\bar{\Gamma}$ represent semiclassical states. It may not.

But clearly this question may gives us some interesting information about loop quantum gravity.
Maybe generalising this toy model
we have presented may help to prove more formally that loop quantum gravity really reproduces
gravity in the semiclassical limit. 

\bigskip

\bigskip

\danger{Acknowledgements} I want to thank Carlo Rovelli for finding time to listen and discuss with me,
some issues concerning the idea presented in this paper. I want to thank him also for
his hospitality at the Centre de Physique Theorique de Luminy where this idea started.


\newpage
  
 

\begin{thebibliography}{99}


\bibitem{r} Carlo Rovelli, Quantum Gravity, Cambridge Monographs on Mathematical Physics, Cambridge
                  University Press, 2004
                  
\bibitem{t} Thomas Thiemann, Modern Canonical Quantum General Relativity, Cambridge Monographs on 
                   Mathematical Physics, Cambridge
                  University Press, 2007              

\bibitem{t2} Thomas Thiemann, Gauge Field Theory Coherent States (GCS) : I. General Properties, 
                      Class.Quant.Grav. \danger{18} (2001) 2025-2064,  arXiv:hep-th/0005233v1

                      

\bibitem{t3} Thomas Thiemann, Complexifier Coherent States for Quantum General Relativity, 
                     Class.Quant.Grav. \danger{23} (2006) 2063-2118, arXiv:gr-qc/0206037v1
      


\bibitem{ac} Emanuele Alesci, Francesco Cianfrani, Quantum Reduced Loop Gravity: Semiclassical limit,
                       Phys. Rev. D \danger{90}, 024006 , 2014,
                        arXiv:1402.3155v1 [gr-qc]


\bibitem{v} Erik Verlinde, On the origin of gravity and the laws of Newton, JHEP 1104: \danger{029}, 2011, 
                    arXiv:1001.0785v1 [hep-th]


\bibitem{s} Lee Smolin, Newtonian gravity in loop quantum gravity,  arXiv:1001.3668v2 [gr-qc]

\bibitem{k} Jerzy Kowalski-Glikman, A note on gravity, entropy, and BF topological field theory, 
                    Phys.Rev. D\danger{81} (2010) 084038, arXiv:1002.1035v1 [hep-th] 


\bibitem{bm} Bojan Mohar, Graph Laplacians, Topics in Algebraic Graph Theory, 
                        Edited by Lowell W. Beineke and Robin J. Wilson, Cambridge University Press, 2007, 
                        113-136


\bibitem{bm2} Bojan Mohar, Some Applications of Laplace Eigenvalues of Graphs, Graph Symmetry: 
                         Algebraic Methods and Applications, Eds. G. Hahn and G. Sabidussi, 
                         NATO ASI Ser. C 497, Kluwer, 1997, pp. 225-275.

\bibitem{rv} Carlo Rovelli, Francesca Vidotto, Single particle in quantum gravity and Braunstein-Ghosh-Severini entropy
                       of a spin network, Physical Review D \danger{81}, 044038 (2010),
                       arXiv:0905.2983v2 [gr-qc]
                       

\bibitem{w} William J. Stewart, Probability, Markov Chains, Queues, and Simulation:
                     The Mathematical Basis of Performance Modeling, Princeton University Press, 2009


\bibitem{o} Olle Haggstrom, Finite Markov Chains and Algorithmic Applications, London Mathematical Society, 
                    Cambridge University Press, 2002


\bibitem{km} Krzysztof A. Meissner, Black hole entropy in Loop Quantum Gravity, Class.Quant.Grav. \danger{21} 
                      (2004) 5245-5252, arXiv:gr-qc/0407052v1


\bibitem{ml} Marcin Domagala, Jerzy Lewandowski, Black hole entropy from Quantum Geometry,
                       Class.Quant.Grav. \danger{21} (2004) 5233-5244, arXiv:gr-qc/0407051v2


\bibitem{gm} A. Ghosh, P. Mitra, An improved estimate of black hole entropy in the quantum geometry approach,
                        Phys.Lett. B \danger{616} (2005) 114-117, arXiv:gr-qc/0411035v3

\end{thebibliography}
\end{document}